\documentclass[preprint,showpacs,preprintnumbers,amsmath,amssymb]{revtex4}
\usepackage{graphicx}% Include figure files
\usepackage{dcolumn}% Align table columns on decimal point
\begin{document}
%\DeclareGraphicsExtensions{.pdf,.gif,.jpg}
\preprint{preprint - vortex group, smc/nims}
\title{History and metastability effects in the intermediate state of mesoscopic type I superconducting Indium}

\author{Ajay D. Thakur$^{1,~\star}$, Ujjal K. Gautam$^2$, Goutam Sheet$^3$, Tomonobu Nakayama$^2$, Yoshio Bando$^2$, Dmitri Goldberg$^2$, Shuuichi Ooi$^1$, Kazuto Hirata$^1$}
\affiliation{$^1$ National Institute for Materials Science, 1-2-1 Sengen, Tsukuba, Ibaraki 305-0047, Japan\\
$^2$ National Institute for Materials Science, 1-1 Namiki, Tsukuba, Ibaraki 305-0047, Japan\\
$^3$ Northwestern University, 2145 Sheridan Road, Evanston, IL 60208, USA}

\date{\today}

\begin{abstract}
We report detailed magnetic measurements on history and metastability effects in the intermediate state of mesoscopic cylinders of type I superconducting Indium. This includes the observation of topological hysteresis with the signature occurrence of different critical fields during flux entry and flux exit. We show the existence of a plethora of metastable configuration and recipes to access them. We also demonstrate the manifestation of superheating and supercooling of superconducting and normal states, respectively across the superconductivity transition.
\end{abstract}

\maketitle

%\section{Introduction}

In the presence of an external magnetic field, a type-I superconductor undergoes a first order transition from a superconducting (SC) to a normal (N) phase \cite{degennes, huebener}. Although the initial and final phases are homogenous, the transition proceeds through the onset of an intermediate state (IS), which essentially consists of an intricate pattern of SC and N domains. Such a modulated phase structure is seen to exist in a wide variety of condensed matter systems, viz., (i) self organized domain patterns in biphasic quasi-2D systems, e.g., ferro- and ferrimagnetic garnets \cite{hubert}, (ii) interface between two immiscible fluids in Hele-Shaw cells \cite{chen}, (iii) Langmuir amphiphilic monolayers at air-water interface \cite{stine, flores}, (iv) chemical reaction diffusion system \cite{lagzi}, (v) nucleonic matter within a neutron star \cite{buckley}, (vi) adsorbates on a metal substrate \cite{plass}, etc. In such systems, the competition and interplay between short-range interactions associated with a positive interfacial energy and the long range magnetostatic/electrostatic/elastic  interactions lead to a spatial modulation in phases \cite{seul}. This lies at the heart of the phenomena of pattern formation. The topology of the IS in type-I superconductors is guided by a competition between the magnetic energy that facilitates the formation of smaller normal domains and the positive surface energy which favors large domains \cite{degennes}. As IS with various configurations of N-SC domains can be achieved by simply varying the magnetic field and temperature, type-I superconductors undoubtedly provide a model experimental system with easily tunable parameters for studying the intricate physics of pattern formation. The emergence of irreversibility during the processes of flux entry and exit in the IS has been an area of active interest recently \cite{menghini1, menghini2, prozorov1, prozorov2, prozorov3, gourdon, jeudy1, jeudy2, velez1}. Magnetic hysteresis in type-I superconductors was historically ascribed to disorder (defects, dislocations and impurities) and edge barriers \cite{livingston}. However, recent developments in high resolution magneto-optical imaging (MOI) techniques have paved way for a revision of archaic ideas. In particular the experimental observations by Menghini {\it et al} \cite{menghini1, menghini2}, Prozorov {\it et al} \cite{prozorov1, prozorov2, prozorov3}, and Gourdon {\it et al} \cite{gourdon} point to the occurrence of topological irreversibility associated with the formation of laminar patterns during flux exit in contrast to the tubular structures formed during flux entry. Furthermore, there has been observation of distinct transition fields during flux entry and flux exit. The behavior is expected to get richer in the mesoscopic regime, where, confinement effects are expected to play an important role \cite{peet}. However, a type-I superconductor has been seen to go into the type-II regime at mesoscopic length scales and this had made it practically impossible to experimentally study the nature of intermediate states in mesoscopic samples. Recently it was demonstrated that pure Indium (In) in the form of nanowires (nanocylinders) encapsulated within ZnS nanotubes remains a type-I superconductor down to diameters smaller than the coherence length ($\xi_{bulk} = 364~nm$) of bulk Indium \cite{gs}. MOI measurements on such In samples are not practicable due to the resolution limit of the garnet films, however, a lot of information regarding the nature of the IS can be obtained by sensitive magnetization studies. In this letter, we demonstrate via magnetic measurements: (i) the occurrence of topological hysteresis in mesoscopic In cylinders and the observation of different critical fields during flux entry and flux exit, (ii) the existence of a plethora of metastable configuration and recipes to access them, and (iii) the superheating and supercooling of SC and N state respectively across the SC-N transition. Based on the results, useful and subtle analogies can be made for other systems manifesting a modulated phase structure due to an interplay of short range and long range interactions.

The sample comprising In-ZnS core-shell structures was synthesized in a high-temperature vertical induction furnace utilizing the recipe described before \cite{ukg}. The charge consisting of 0.1 g of In$_2$S$_3$, 0.25 g of ZnS, and 0.05 g of activated carbon was thoroughly mixed and put in a graphite crucible placed at the center of the furnace. An additional 0.5 g of S was put in another graphite crucible that was placed at the bottom of the furnace. Initially the furnace was evacuated to 1$\times$10$^{-5}$ Pa and then filled with 99.99 $\%$ of Argon gas. The crucible was rapidly heated to 1300$^oC$, where it was kept for 45 min in flowing (0.15 l/min) Ar gas and then the furnace was switched off to allow a natural cooling to room temperature. Sample was collected from several locations from the deposition tube placed at the top of the crucible. High resolution transmission electron microscopy (TEM) and energy dispersive spectroscopy (EDS) mapping on the sample was performed using a JEOL JEM-3000F (300kV) TEM. For this, the core-shell structures were first dispersed in ethanol by ultrasonication and then deposited dropwise onto a holey carbon coated copper TEM grid. The magnetic measurements were performed using a Quantum Design MPMS XL magnetometer utilizing the reciprocating sample option (RSO) with 1 cm amplitude. The MPMS XL magnet has a field uniformity of 0.01 $\%$ over 4 cm length leading to a maximum field inhomogeneity of 0.02 Oe during the RSO measurements in our experiments. Ac susceptibility ($\chi_{ac}^{\prime}$) measurement was performed using ac-option of Quantum Design MPMS SQUID magnetometer.

Figure 1 shows the $\chi_{ac}^{\prime}$ versus $T$ plot for a collection of In-ZnS core-shell structures obtained at an ac-drive ($H_{ac}$) of 1 Oe with drive frequencies ($f$) of 1 Hz, 11 Hz, and 111 Hz, respectively (shown by various open symbols). The onset of superconducting transition is marked by $T_c^{on}$ and is independent of the drive frequencies in the range 1$\mid$111 Hz. In the main inset is shown the TEM image of a single core-shell structure comprising of a ZnS shell encapsulating an In core. The sub-inset panels (a) to (c) shows the EDS maps of the core-shell structure corresponding to the In L-edge, Zn K-edge and S K-edge spectra respectively. The x-ray diffraction for the as prepared samples and the selective area electron diffraction (SAED) of an individual core-shell structure are reported elsewhere \cite{ukg}. They attest to the single-crystalline nature of the In core. The distribution of length and diameters of the final core-shell structure are sensitive to the exact synthesis conditions. For the sample used in the present study, the average external diameter of the core-shell structure was about 200 nm, with an average diameter of 120 nm for the In nanowire core. The nanowires had a typical length of 5-20 $\mu$m. Using these quantities, an estimate for the amount of Indium present in the sample was done for the purpose of normalizing the magnetization data. For this purpose, the following expression was used:
\begin{equation}
M_{In} = \frac{M_{sample}}{\left[ 1 + \left( \frac{\langle D_{ZnS} \rangle^2 - \langle D_{In} \rangle^2}{\langle D_{In} \rangle^2} \right) \frac{\rho_{ZnS}}{\rho_{In}}\right]}
\end{equation}
where, $M_{In}$ is the estimated weight of Indium present in collection of core-shell structures with a total weight $M_{sample}$, $\langle D_{ZnS}\rangle$ and $\langle D_{In} \rangle$ are the average diameters of the ZnS shell and Indium core respectively. Also, $\rho_{ZnS}$ (4.09 gm/cm$^3$) and $\rho_{In}$ (7.31 g/cm$^3$) are the densities of ZnS and Indium, respectively. Different weights of sample (in the range of 2 to 4 mg) were measured to compare the normalized moment under identical conditions and the estimated normalized moments were identical within experimental resolution suggesting that the core-shell structures are randomly oriented with respect to the field direction. Figure 2 shows the magnetization hysteresis ($M-H$) loops at 2.5 K in the case of a collection of In nanowires (panel (a)), and bulk In cylinder with a diameter 0.9 mm and length 2.1 mm and field perpendicular to the axis of the cylinder (panel (b)). Inset in panel (b) shows an expanded view of a portion of the $M-H$ loop. The notional critical fields, $H_{c}^{for}$ and $H_{c}^{ret}$, for the forward and the return legs of the $M-H$ loop, respectively are marked. As can be observed in the nanowire case, the critical fields during the field sweep-up and field sweep-down are markedly different. Also, the critical field in the case of the bulk cylinder has a value which lies between $H_{c}^{for}$ and $H_{c}^{ret}$ for the nanowire case. This is remarkably, in contrast to the situation in the case of nanowires of superconducting Pb (reported as a type-II superconductor in the nanowire form) \cite{dai}, where there is a unique critical field whose value is markedly higher than the value of $H_c$ for bulk Pb. This attests to the type-I nature of the In nanowires. The differences between the critical fields during the field sweep-up and field sweep-down can be attributed to the phenomenon of superheating and supercooling where the ordered superconducting (disordered normal) phase is superheated (supercooled) across the first order SC-N phase transition. In such a scenario, there is the possibility for the existence of a multiplicity of superheated and supercooled metastable states. The nature of these metastable states depends on the exact magneto-thermal history during their formation. In order to elucidate the phenomena of superheating and supercooling across the SC-N transition in the case of In nanowires and for accessing and exploring the various metastable configurations, a set of protocols were followed to prepare the state of the system for magnetic measurements. These are enumerated below:

\renewcommand{\theenumi}{\Roman{enumi}}
\begin{enumerate}
\item Field cooling followed by a temperature shake (FCTS) : the sample is initially exposed to an external field, $H_{FC}$ at $T_{IN}>T_c$ and then cooled to a temperature $T_{FC}<T_c$ leading to the field cooled state (FC). This FC state is then taken to a temperature $T_{FC} + \Delta T_{SH}$ and $T_{FC} - \Delta T_{SH}$ for the positive and negative temperature shake, respectively and then brought back to $T_{FC}$ before measurement.
\item Field cooling followed by a field shake (FCFS) : after preparing the FC state, the sample is measured while sweeping the field up and down by an amount $\Delta H$ about $H_{FC}$ with the final measurement done at $H_{FC}$ after equal number of field sweeps in both directions.
\item Field shake at a given field on the envelope $M-H$ loop (FSL) : at various locations on the envelope $M-H$ loop ($H_X$, $T_X$), minor hysteresis loops are made by sweeping the field up and down by an amount $\Delta H$ and after equal number of field sweeps in both directions, the sample is brought back to the starting location ($H_X$, $T_X$). 
\end{enumerate}   
Figure 3 shows a portion of the magnetization hysteresis ($M-H$) loop at 2.5 K (black line and symbols) with arrows marking the field sweep directions. The red diamond symbols corresponds to the magnetization values for the various metastable states accessed via the FCTS protocol (Protocol-I). The inset shows a color scale contour plot for the evolution of magnetization values from the field cooled state for various negative temperature shakes. There exists a considerable similarity between the distribution of various metastable states (shown by the red diamond symbols in Fig.3) observed experimentally in the present case and the theoretical calculations recently demonstrated by Peeters et al \cite{peet}. In particular, the near absence of metastable configurations at the low field and the high field end indeed is remarkable. 

Figure 4 shows a portion of the $M-H$ loop at 2.5 K (black solid symbols and curve) with arrows indicating the field sweep directions. In violet (open symbols and curve) and red (open symbols and curve) are shown typical measurements following the FCFS protocol (Protocol-II), with I1 and I2 (F1 and F2) denoting the respective initial (final) states. As can be observed, beyond a certain number of field shakes, further changes in the M values is notionally absent. Thus, F1 and F2 respectively indicate the limiting M values at the given field values. Figure 5 shows a portion of the magnetization hysteresis ($M-H$) loop at 2.5 K (black line) with arrows marking the field sweep directions. Also shown are various minor loops obtained via the FSL protocol (Protocol-III). The dotted arrows mark the evolution from the initial to the final states (different open symbols at various field values) following the above protocol. The inset shows a plot of changes in magnetization values via the FSL protocol at various field values. As can be clearly seen, the difference between the initial FC state and the final state obtained as a result of the subsequent field shake, peaks at intermediate field values. This reinforces the observations of Fig.3, that the metastability is maximal at intermediate field values in agreement with the theory \cite{peet}. Panel (a) of Fig. 6 shows a color scale contour plot of the evolution of the magnetic susceptibility values from the field cooled state following the FCTS protocol with negative temperature shakes obtained at 2.5 K. In panel (b) is shown a similar plot as panel (a) above, however, positive temperature shakes are employed in this case and the starting configuration corresponds to the final state in panel (a). Panels (a) and (b) respectively portray the hallmark of the phenomena of superheating and supercooling. As can be seen clearly, whereas in panel (a), a low temperature superconducting phase is superheated, in panel (b), a high field normal state is supercooled.

A number of different factors contribute to the observed hysteresis in a typical M-H measurements, viz., (a) pinning, (b) geometrical barrier, (c) topological, (d) superheating and supercooling across the first order transition, etc. If the hysteresis is of a pinning induced nature, it is expected to increase with a decrease in field and attains a maximum at zero field. This is not what is observed in the present case. A geometrical barrier driven hysteresis is unlikely to contribute due to the large aspect ratio for the In cylinders in the mesoscopic regime. The differences the topology of normal domains during flux entry (tubular) and flux exit (lamellar) leads to topological hysteresis. In macroscopic samples, the occurence of topological hysteresis was notionally understood by assuming the presence of a surface barrier during flux entry, which was absent during flux exit such that a large number of laminae are connected to the sample edge during flux exit leading to a continuous flux exit. in contrast during flux entry, normal domains are pulverized into smaller pieces during flux entry. Recent MOI measurements lead to a more refined understanding of the phenomena. Recent theoretical work by Berdiyorov et al suggested the possibility of a confinement ( topology) enriched hysteresis with the possibility of the existence of a plethora of metastable states. Our observations appear to be in good agreement with their proposal. In addition we observe signatures of superheating and supercooling across the N-SC transition in the mesoscopic regime. There exists ample evidence in the literature for superheating/supercooling induced asymmetry in M-H measurements with accompanying hysteresis in the case of first order order-disorder transition across peak effect phenomena in type-II superconductors. An analogy can be drawn to explain the observed asymmetry in the present case. Due to the absence of directionality with respect to the applied field in the present case for the nanowires, the demagnetization correction is not possible, however such a correction is just expected to renormalize the observed $M$ values. The present results on hysteresis and the accompanying metastability effects in the intermediate state of mesoscopic type-I superconducting Indium attempts to provide a glimpse into the richness of the phenomena. Based on the observations of superheating and supercooling in the intermediate state across the SC-N transition, interesting physical analogy could be made for other systems where the competition and interplay between short-range interactions associated with a positive interfacial energy and the long range magnetostatic/electrostatic/elastic  interactions lead to a spatial modulation in phases, e.g., by tuning suitable parameters, the rate of forward (backward) reaction could possibly be modified in a typical reaction-diffusion system.

In conclusion, we report the observation of topological hysteresis in a collection of mesoscopic cylinders of type-I superconducting In. We demonstrate the occurence of a plethora of metastable magnetic configurations, depending on the exact recipe followed to obtain a particular configuration. Also reported is the occurence of the phenomena of superheating and supercooling across the imminent first order SC-N phase transformation.

$^{\star}$~~thakur.ajay@nims.go.jp

\newpage

\begin{figure}[!h]
\includegraphics[scale=0.6,angle=0]{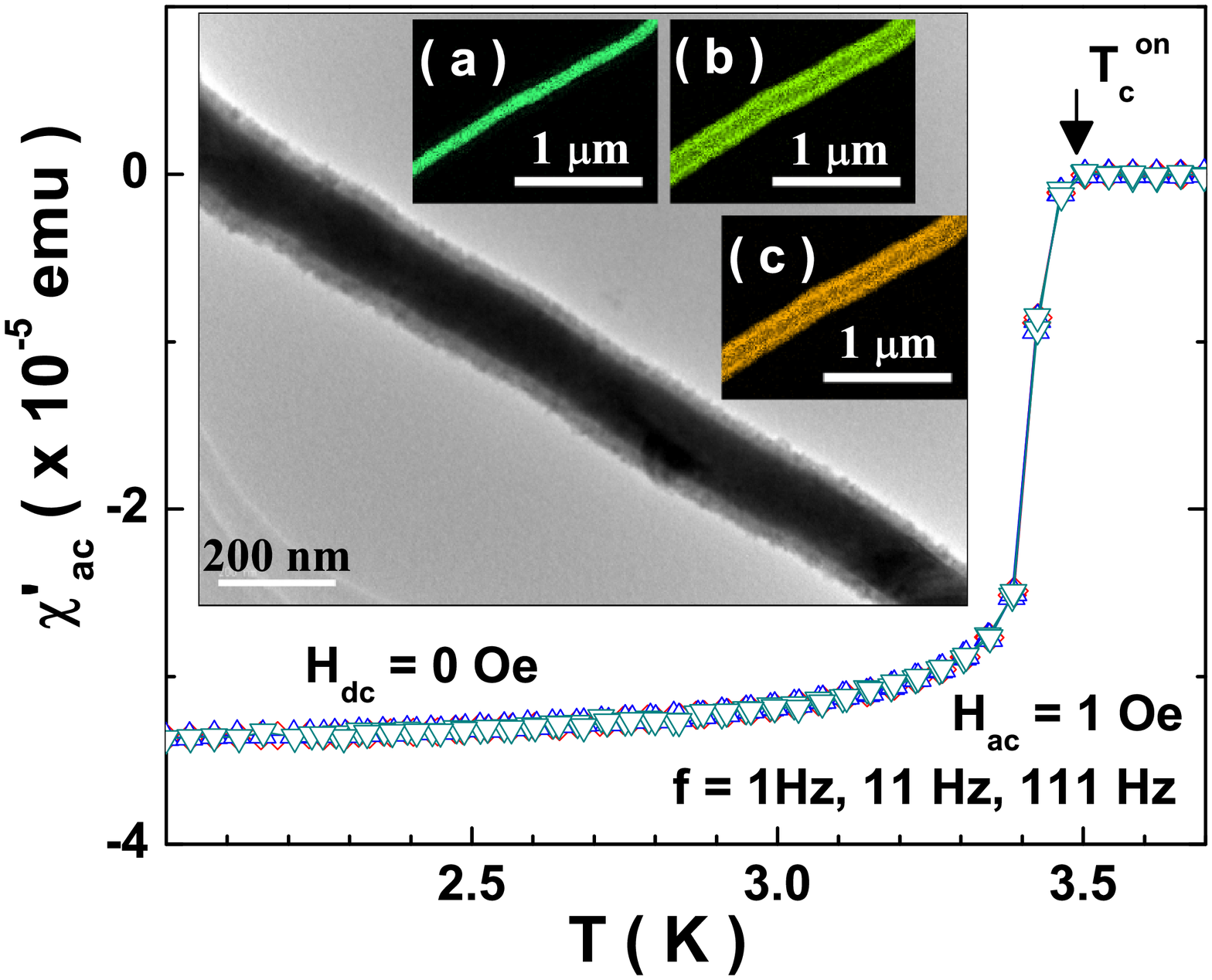}
\caption{(Color online) $\chi_{ac}^{\prime}$ versus $T$ plot obtained at drive frequencies of 1 Hz, 11 Hz and 111 Hz (various open symbols). The onset of superconducting transition is marked by $T_c^{on}$. The main inset shows the TEM image of a single core-shell structure comprising of a ZnS shell encapsulating an In core. The sub-inset panels (a) to (c) shows the EDS maps of the core-shell structure corresponding to the In L-edge, Zn K-edge and S K-edge spectra respectively. }
\end{figure}

\begin{figure}[!h]
\includegraphics[scale=0.6,angle=0]{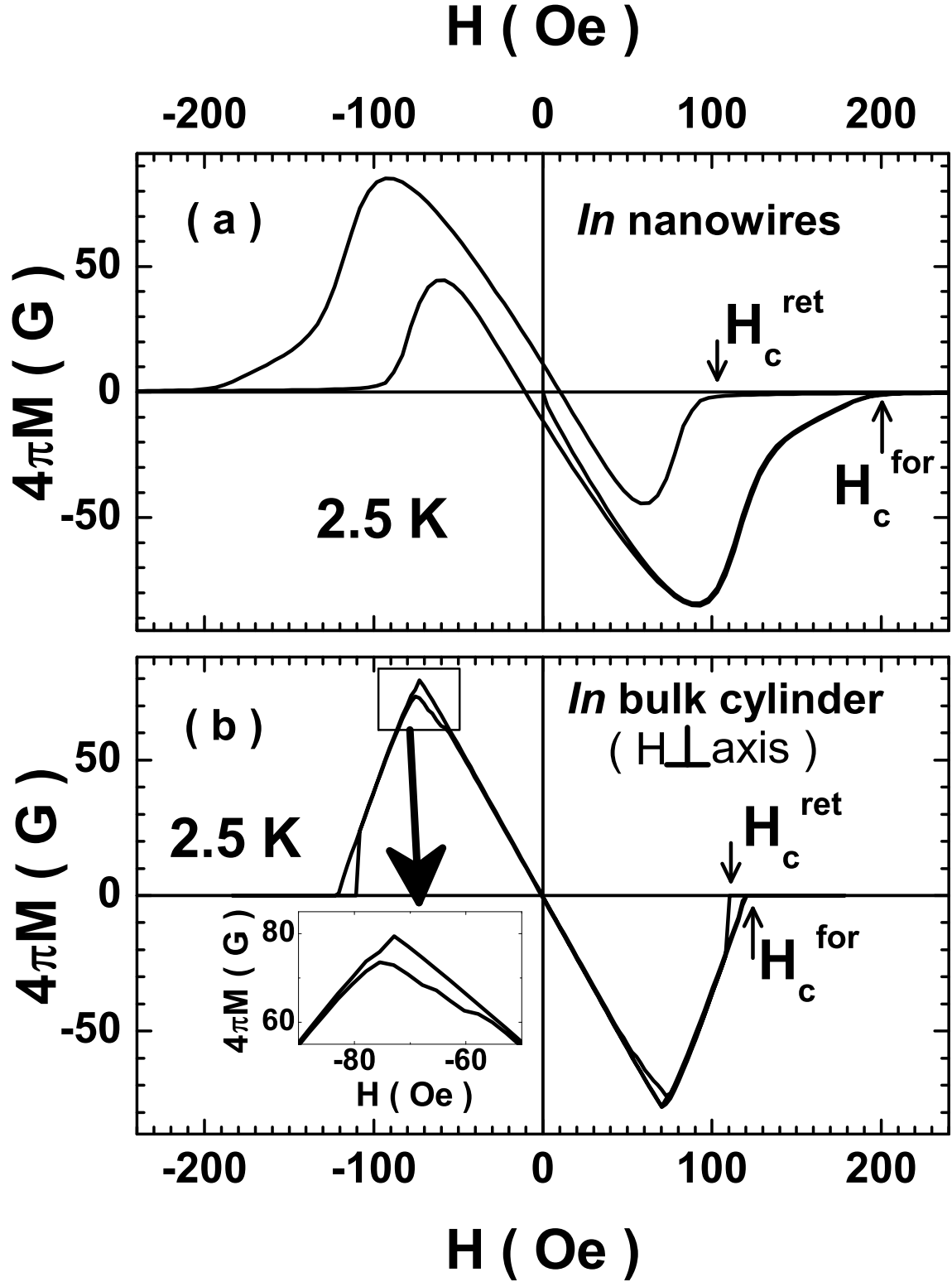}
\caption{(Color online) Magnetization hysteresis ($M-H$) loops at 2.5 K in the case of (a) a collection of In nanowires, and (b) bulk In cylinder with a diameter 0.9 mm and length 2.1 mm ($H\perp$ axis).  The critical fields $H_c^{for}$ and $H_c^{ret}$ during the forward and return legs of the $M-H$ loop, respectively are labelled.}

%The inset in panel (a) shows the $M-T$ curve in the case of encapsulated In nanowires. The corresponding $M-T$ curve for the bulk In cylinder is shown as inset (i) in panel (b). Inset (ii) in panel (b) shows an expanded view of a portion of the $M-H$ loop and the inset (iii) shows the $M-H$ loop for the bulk In cylinder with field applied parallel to its axis. The transition temperatures ($T_c$) for both the cases are marked and also labelled are the notional critical fields, $H_{cf}$ and $H_{cr}$, for the forward and the return legs of the $M-H$ loop, respectively.
\end{figure}

\begin{figure}[!h]
\includegraphics[scale=0.6,angle=0]{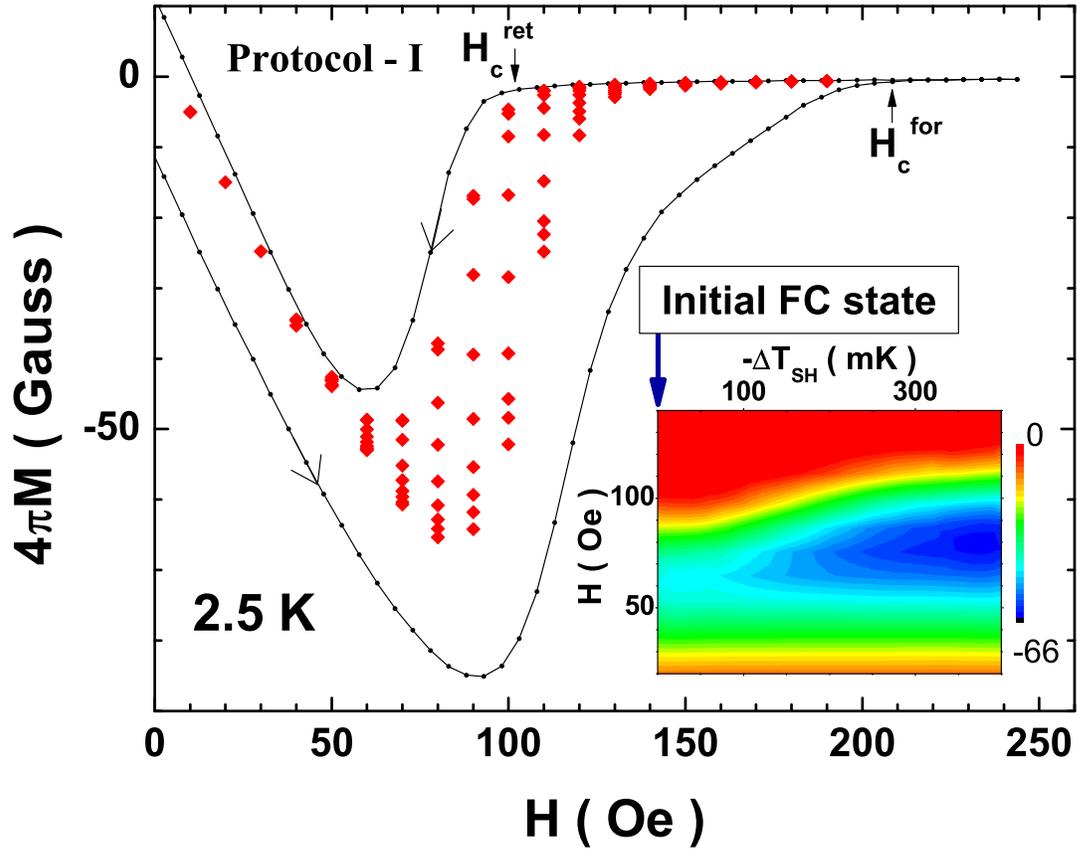}
\caption{(Color online) A portion of the magnetization hysteresis ($M-H$) loop at 2.5 K (black line and symbols) with arrows marking the field sweep directions. The red diamond symbols corresponds to the magnetization values for the various metastable states accessed via the FCTS protocol (see text for details). The inset shows a color scale contour plot for the evolution of magnetization values from the field cooled state for various temperature shakes.}
\end{figure}

\begin{figure}[!h]
\includegraphics[scale=0.6,angle=0]{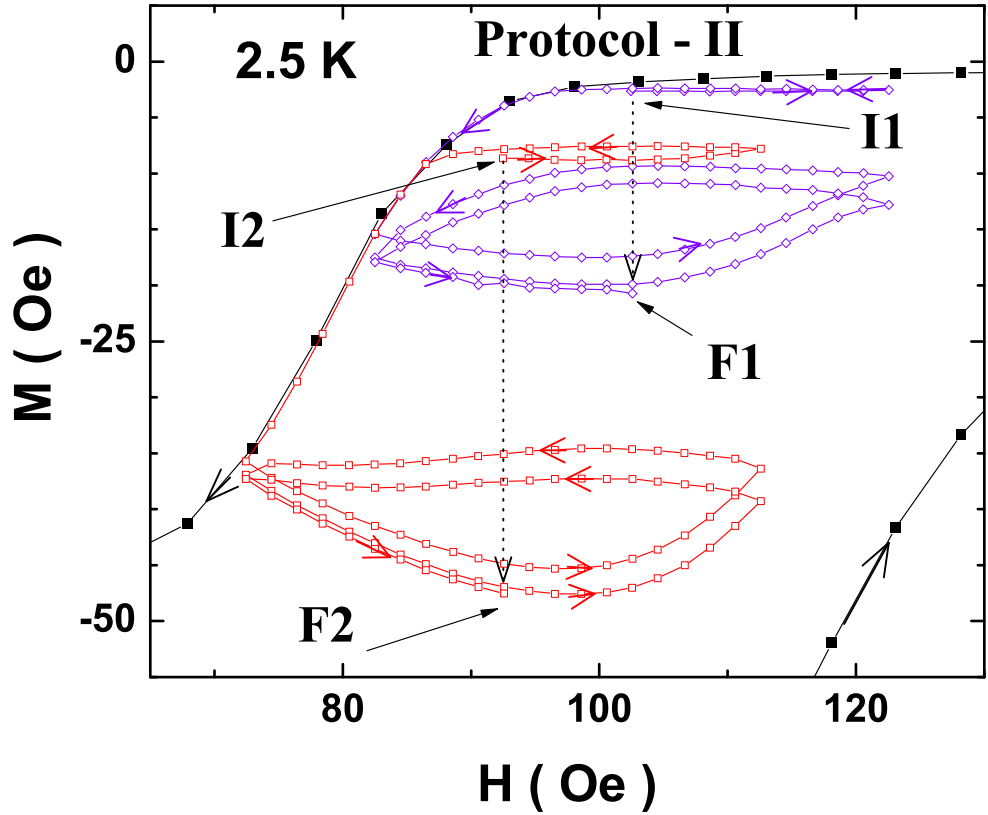}
\caption{(Color online) Portion of the M-H loop at 2.5 K (black solid symbols and curve) with arrows indicating the field sweep directions. In violet (open symbols and curve) and red (open symbols and curve) are shown typical measurements following the FCFS protocol (see text). I1 and I2 (F1 and F2) denote the respective initial (final) states}
\end{figure}

\begin{figure}[!h]
\includegraphics[scale=0.6,angle=0]{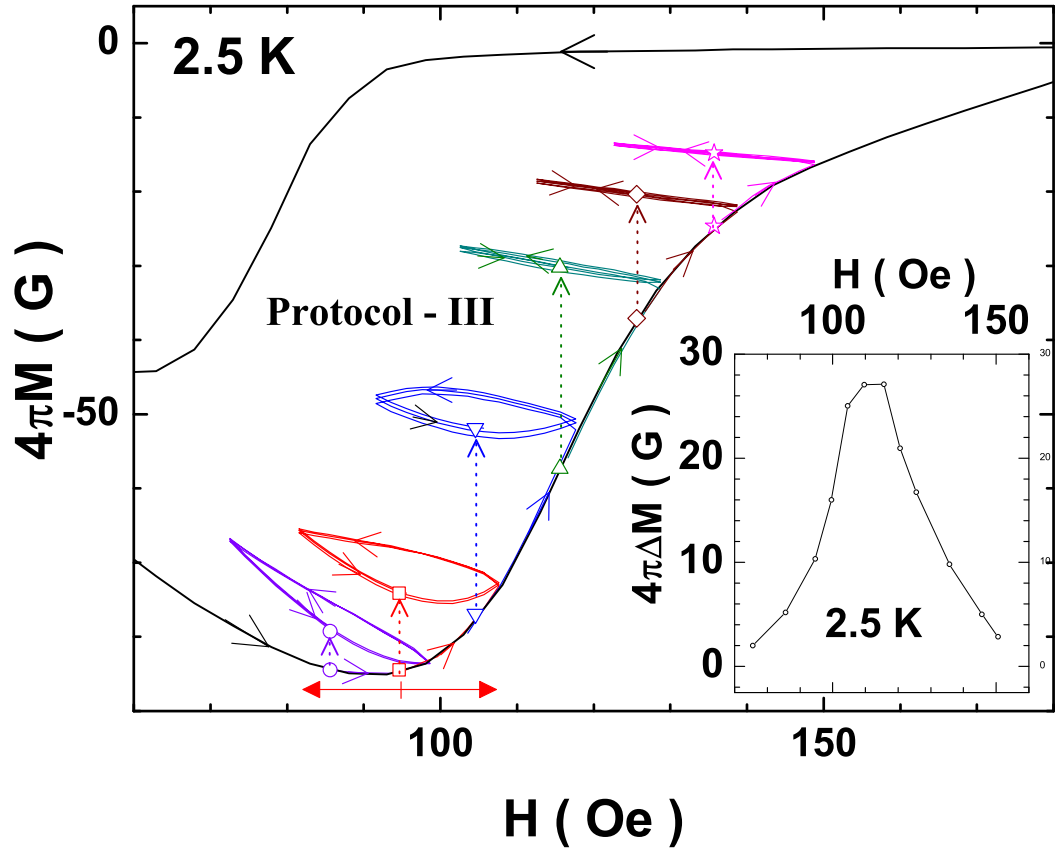}
\caption{(Color online) A portion of the magnetization hysteresis ($M-H$) loop at 2.5 K (black line) with arrows marking the field sweep directions. Also shown are various minor loops obtained via the FSL protocol (see text for further details). The dotted arrows mark the evolution from the initial to the final states (different open symbols at various field values) following the above protocol. The inset shows a plot of changes in magnetization values via the FSL protocol (see text) at various field values.}
\end{figure}

\begin{figure}[!h]
\includegraphics[scale=0.6,angle=0]{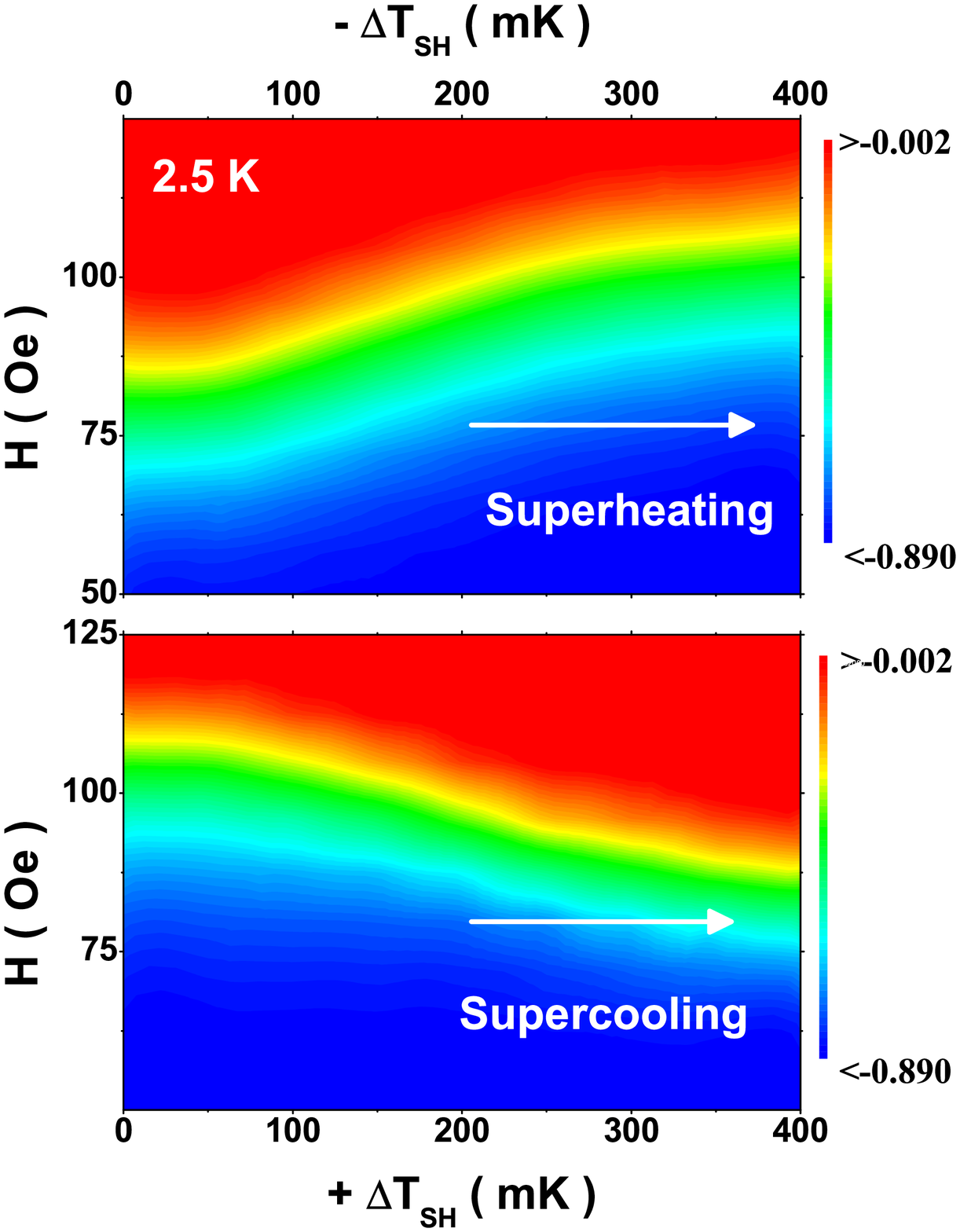}
\caption{(Color online) (a) Color scale contour plot of the evolution of the magnetic susceptibility values from the field cooled state following the FCTS protocol (see text) with negative temperature shakes (see text for details) obtained at 2.5 K. (b) A similar plot as (a) above, however, positive temperature shakes are employed in this case and the starting configuration corresponds to the final state in (a) above.}
\end{figure}

\begin{references}
\bibitem{degennes}
P. G. de Gennes, {\it Superconductivity of Metals and Alloys} (W. A. Benjamin, Inc., New York, 1966).

\bibitem{huebener}
R. P. Huebener, {\it Magnetic Flux Stuctures in Superconductors} (Springer-Verlag, Heidelberg, 1979).

\bibitem{hubert}
A. Hubert and R. Sch$\ddot{a}$fer, {\it Magnetic Domains} (Springer, Berlin, 2000).

\bibitem{chen}
J. -D. Chen, Experiments in fluids 5, 363 (1987).

\bibitem{stine}
K. J. Stine, Ch. M. Knobler, and R. C. Desai, Phys. Rev. Lett. 65, 1004 (1990).

\bibitem{flores}
A. Flores, E. Corvera-Poir\'{e}, C. Garza, R. Castillo, J. Phys. Chem. B 110, 4824 (2006).

\bibitem{lagzi}
I. Lagzi and D. Ueyama, Chemical Physics Letters 468, 188 (2009).

\bibitem{buckley}
K. B. W. Buckley, M. A. Metlitski, and A. R. Zhitnitsky, Phys. Rev. Lett. 92, 151102 (2004).

\bibitem{plass}
R. Plass, N. C. Bartelt, and G. L. Kellogg, J. Phys.: Cond. Matt. 14, 4227 (2002).

\bibitem{seul}
M. Seul and D. Andelman, Science 267, 476 (1995).

\bibitem{prozorov1}
R. Prozorov, R. W. Giannetta, A. A. Polyanskii, and G. K. Perkins, Phys. Rev. B72, 212508 (2005).

\bibitem{prozorov2}
R. Prozorov, Phys. Rev. Lett. 98, 257001 (2007).

\bibitem{prozorov3}
R. Prozorov, A. F. Fidler, J. R. Holberg, and P. C. Canfield, Nat. Phys. 4, 327 (2008).
 
\bibitem{menghini1} 
M. Menghini and R. J. Wijngaarden, Phys. Rev. B 72, 172503 (2005). 

\bibitem{menghini2}
M. Menghini and R. J. Wijngaarden, Phys. Rev. B 75, 014529 (2007).

\bibitem{gourdon}
C. Gourdon, V. Jeudy, and A. Cebers, Phys. Rev. Lett. 96, 087002 (2006).

\bibitem{jeudy1}
V. Jeudy, C. Gourdon, A. C$\bar{e}$bers, and T. Okada, J. Appl. Phys. 101, 09G118 (2007).

\bibitem{jeudy2}
V. Jeudy, C. Gourdon, A. C$\bar{e}$bers, and T. Okada, Phys. Rev. B 72, 014513 (2005).

\bibitem{velez1}
Sa$\ddot{u}$l V$\acute{e}$lez, Carles Panad$\grave{e}$s-Guinart, Guillem Abril, Antoni Garc'a-Santiago, Joan Manel Hernandez, and Javier Tejada, Phys. Rev. B 78, 134501 (2008).

\bibitem{livingston}
J. D. Livingston and W. DeSorbo, in Superconductivity, edited by R. D. Parks (Dekker, New York, 1969), Vol. 2, p. 1235.

\bibitem{peet}
G. R. Berdiyorov, A. D. Hernandez, and F. M. Peeters, Phys. Rev. Lett. 103, 267002 (2009).

\bibitem{dai}
X. Y. Zhang and J. Y. Dai, Nanotechnology 15, 1166 (2004).

\bibitem{gs}
G. Sheet, U. K. Gautam, A. D. Thakur, K. Hirata, Y. Bando, and T. Nakayama, Appl. Phys. Lett. 94, 053108 (2009).

\bibitem{ukg}
U. K. Gautam, X. Fang, Y. Bando, J. Zhan, and D. Goldberg, ACS Nano 2, 1015 (2008).





\end{references}
\end{document}